\documentclass[letterpaper,superscriptaddress,aps,pra,nolongbibliography,twocolumn,showpacs,floatfix]{revtex4-1} % {{{
\usepackage{graphicx,helvet}
\usepackage{color}
\usepackage{bm}                   
\usepackage{tikz}
\usepackage{amsfonts}
\usepackage{lipsum}
\definecolor{mygray}{gray}{0.4}
\definecolor{light-blue}{rgb}{0.38,0.58,0.69}
\definecolor{myred}{rgb}{0.83,0.34,0.37}
\definecolor{mygreen}{rgb}{ .59,.74,.61}
\definecolor{miviol}{rgb}{.57,.11,.81} %%{0.55,0.48,0.80}
\usepackage{amsmath}%
\usepackage{MnSymbol}%
\usepackage{wasysym}%
 \usepackage{soul}
\usetikzlibrary{decorations.shapes}

\newcommand{\beqa}{\begin{eqnarray}}
\newcommand{\eeqa}{\end{eqnarray}}
\newcommand{\beq}{\begin{equation}}
\newcommand{\eeq}{\end{equation}}
\newcommand{\ktimes}{\rangle\! \langle}
\newcommand{\op}[2]{|#1\ktimes #2|}
\usepackage{amsmath}

\newcommand{\Eref}[1]{Eq.~(\ref{#1})}

\newcommand{\tr}{\mathop{\mathrm{Tr}}\nolimits}

\newcommand{\ifimar}{Instituto de Investigaciones Físicas de Mar del Plata (IFIMAR), Facultad de Ciencias Exactas y Naturales, Universidad Nacional de Mar del Plata \& CONICET, 7600 Mar del Plata, Argentina.}
\newcommand{\conicet}{Consejo Nacional de Investigaciones Cient\'ificas y Tecnol\'ogicas (CONICET), Buenos Aires, Argentina}
\newcommand{\uba}{Departamento de F\'isica ``J. J. Giambiagi'' and IFIBA, FCEyN, Universidad de Buenos Aires, 1428 Buenos Aires, Argentina}
\newcommand{\ipcms}{\mbox{Universit\'e de Strasbourg, CNRS, Institut de Physique et Chimie des Mat\'eriaux 
de Strasbourg,} UMR 7504, F-67000 Strasbourg, France}
\newcommand{\cnea}{Departamento de F\'isica Te\'orica, Comisi\'on Nacional de Energ\'ia At\'omica, Buenos Aires, Argentina}
\newcommand{\unsam}{Escuela de Ciencia y Tecnolog\'ia, Universidad Nacional de San Mart\'in, San Mart\'in, Argentina}

\newcommand{\hF}{\hat{F}}
\newcommand{\hT}{\hat{T}}
\newcommand{\hX}{\hat{X}}
\newcommand{\hP}{\hat{P}}
\newcommand{\hU}{\hat{U}}
\newcommand{\hV}{\hat{V}}
\newcommand{\hA}{\hat{A}}
\newcommand{\hB}{\hat{B}}
\newcommand{\su}[1]{\text{\textbf{\textsf{#1}}}}        %
\newcommand{\s}{\su{S}}                                 %
\newcommand{\se}{\s_\epsilon}
\begin{document}
\title{Chaos Signatures in the Short and Long time Behavior of the Out-of-time Ordered Correlator}
\author{Ignacio Garc\'ia-Mata} \affiliation{\ifimar} \affiliation{\conicet}
\author{Marcos Saraceno}\affiliation{\cnea}\affiliation{\unsam}
\author{Rodolfo A. Jalabert} \affiliation{\ipcms} 
\author{Augusto J. Roncaglia}\affiliation{\uba}
\author{Diego A. Wisniacki} \affiliation{\uba}
% \email{@gmail.com}
\begin{abstract}
Two properties are needed for a classical system to be chaotic: exponential stretching and mixing.
Recently, out-of-time order correlators  were proposed as a measure of chaos in a wide range of physical 
systems. While most of the attention has previously been devoted to the short time stretching aspect of chaos, characterized by the Lyapunov exponent, we show for quantum maps that the out-of-time correlator approaches its stationary value exponentially with a rate determined by the Ruelle-Pollicot resonances. 
This property constitutes clear evidence of the dual role of the underlying classical chaos dictating the behavior of the correlator at different timescales.
\end{abstract}
%
%03.65.Yz	Decoherence; open systems; quantum statistical methods
%--
%03.65.Ta  "Foundations of quantum mechanics"
%--
%03.67.-a	Quantum information (see also 42.50.Dv Quantum state
% engineering and measurements; 42.50.Ex Optical implementations
% of quantum information processing and transfer in quantum optics)
%--
%05.45.Mt	Quantum chaos; semiclassical methods
%%----------------------------------
%
\date{12 June 2018}

\maketitle

\noindent
%%%%%Intro
\textit{Introduction.} Classical chaos is commonly associated with the
exponential separation of trajectories. The impossibility  of translating this concept literally into the quantum realm is one of the reasons why the development of quantum chaos
\cite{haake,*stockmann} has remained a challenge throughout the years. Nevertheless, important achievements have been accomplished in the field; one of the most prominent being the description of universal properties in the spectrum described by ensembles of random matrices. Thus, spectral properties can in many cases signal quantum chaos. 
More recently, tools like the Loschmidt echo \cite{DiegoScholar} have allowed to gauge and analyze quantum chaos in the time domain. 
Chaotic (ergodic) properties of eigenfunctions have become a central subject in relation to thermalization (via the eigenstate thermalization hypothesis \cite{Jensen85,*Deutsch91,*Srednicki1994,*Rigol2008}), particularly in  many-body localization \cite{nandkishore2015}.

Another probe of quantum chaos is provided by  
the out-of-time ordered correlator (OTOC). 
It was first used in Ref.~\cite{larkin1969quasiclassical}, where its exponential growth with time 
was associated to chaotic behavior.
Recently the OTOC has been put forward  
as a measure of chaos in many-body systems 
\cite{shenker2014black,kitaev2015simple,shenker2015stringy,aleiner2016,huang2017,chen2017,tsuji2017,kurchan2016,borgonovi2018,bentsen2018fast}. 
The physical concepts behind the OTOC are particularly interesting in that they 
can be related to scrambling of quantum information \cite{campisi2017,Bohrdt2017,Fazio2018} and entanglement.
Moreover, the subject has attracted considerable attention following
the conjecture that puts a bound on its growth rate for many-body thermal quantum systems \cite{maldacena2016}. 
The ever growing advances in coherent manipulation of quantum systems have also permitted to envisage and carry out experiments to measure them \cite{kurchan2016,Swingle2016,campisi2017,Li2017,Garttner2017,Landsman2018}.

These time domain features of the OTOC concern relatively short times (i.e. up to the Ehrenfest time).
General chaotic behavior is characterized by two properties. The first one is stretching, causing exponential separation of trajectories  
and is quantified by the Lyapunov exponent. 
But small time separation of initial conditions is not enough for chaos. 
The second property associated with chaos is mixing, which in a compact phase space is realized when the stretching trajectories fold back unto themselves (simple graphical examples are given by the baker's map or Smale's horseshoe). Mixing takes place for longer times and is quantified by the decay of correlation functions. For strongly chaotic systems this decay is exponential 
with a rate given by Ruelle-Pollicott  resonances (RPRs) \cite{ruelle1986,*ruelle1987}. 
Since the OTOC is by definition a correlation function, an observable effect of this
second regime due to mixing in chaotic systems is to be expected \cite{polchinski2015}.
Such a regime has not to our knowledge been observed in any system for which the OTOC has been studied.

In this Letter we show that the RPRs play an important role in the time behavior of the OTOC. For strongly chaotic systems, the approach to saturation is exponential
and dominated by the largest RPR.
Moreover we provide an analytical proof, for a simple model, that the short time exponential growth is governed by the Lyapunov exponent. Up to now, such behavior has only been shown numerically
or deduced under certain approximations \cite{kurchan2016,Rozenbaum2017,chen2018,rammensee2018,Jalabert2018}.

\noindent
\textit{Out-of-time ordered correlator.}
The OTOC is defined as the thermal 
average involving the %
commutator between two operators at different times
\begin{equation}
\label{otocdef1}
C(t)=\left \langle [\hA(t),\hB][\hA(t),\hB]^{\dagger}\right \rangle 
\end{equation}
with $\hA(t)$ the Heisenberg evolution of the operator $\hA$. For simplicity, we consider operators acting on a Hilbert space of dimension $N$. As the evolution will be governed by a $N \times N$ unitary map 
the average is given by $ \langle .\rangle= \tr (.)/N$. For $\hA$ and $\hB$ Hermitian we can expand the commutators and obtain $C(t)=-2 [O_1(t)-O_2(t)]/N$ with 
\begin{equation}
\label{eqO1}
O_1(t)=\tr \left[\hA(t) \hB \hA(t) \hB\right] , \quad O_2(t)=\tr \left[\hA^2(t) \hB^2\right].
\end{equation}

For the time evolution we use  the quantum version of classical maps on the torus. They embody, in the simplest possible way, the essential features of chaotic motion: 
a compact phase space and an exponential divergence of trajectories,
together with computable Lyapunov exponents, Ruelle-Pollicott resonances, and eventually other quantities that characterize their chaotic nature.  

For concreteness, we use the quantum version of a paradigmatic example of chaos, the perturbed Arnold's cat map \cite{arnold} on the unit torus
\begin{equation}
\begin{array}{lll}
p'&=& p+q - 2 \pi k \sin [2 \pi q] \\
q'&=& q+p'+ 2 \pi k \sin [2 \pi p']	
\end{array}
\ \ {\rm mod}\ 1.
\end{equation}
The Lyapunov exponent for $k=0$ is the logarithm of the largest eigenvalue of 
$M=\begin{pmatrix}
2&1\\
1&1
\end{pmatrix}$, i.e. $\lambda_{\rm L}=\ln[(3+\sqrt{5})/2]$ and does not change significantly for small values of $k$. 
We introduce a nonlinear perturbation to avoid nongeneric behavior that can appear in the quantized version at long times.  

A quantum map is a unitary operator that represents the  canonical transformation corresponding to the classical map. The torus structure implies a double periodicity, 
which upon quantization imposes the discreteness of Hilbert space with an effective Planck constant related to the dimension $N$ of Hilbert space by
$h_{\rm eff}=1/(2\pi N)$. 
The map is quantized as a $N \times N$  unitary operator $\hU_M$ \cite{Hannay1980}, and the time 
evolution is given in discrete steps by  $\hU_M^t$. For convenience, 
we  express it as a composition of ``kicks'' in position and momentum 
\begin{equation}
\label{qcatmap}
\hU_M=e^{-i 2\pi  (\frac{p^2}{2N}- k N\cos(2 \pi p/N))}e^{-i 2 \pi (\frac{q^2}{2N}+k N\cos(2\pi q/N))}
\end{equation}
with $q,p=0,...,N-1$. Position and momentum are related by the discrete Fourier transform. Thus, $\hU_M$ can be efficiently implemented using fast Fourier transform routines.
%%%

An intuitive interpretation of the relation between $C(t)$ and chaos can be given directly 
for position and momentum operators $\hX$ and $\hP$ \cite{maldacena2016}. On the quantized torus, 
these operators are not well defined, but an approximation in the classical limit can be constructed in terms of Schwinger shift operators\cite{schwinger}. These are defined as
\begin{align}
\label{shift}
\hat{V}=\sum_{q\in\mathbb{Z}_N}\op{q+1}{q}\ ;{ }&\ 
\hat{U}=\sum_{q\in\mathbb{Z}_N}\op{q}{q}\tau^{2 q}
\end{align}
with $\tau = e^{i\pi/N}$. 
Then, position and momentum can  be defined as
\begin{align}
\label{XP}
\hX=\frac{\hU-\hU ^\dagger}{2 i}\ ;{ }&\ 
\hP=\frac{\hV-\hV^\dagger}{2 i}
\end{align}
which are Hermitian and in the semiclassical limit fulfill the correct commutation relation.
We proceed to compute  $C(t)$ numerically  replacing $\hA \leftrightarrow \hX$ and $\hB \leftrightarrow \hP$.
%%%%%%%%%%%%%%%%%%%%%%%%%%%%%FIGURA%%%%%%%%%%%%%%%%%%%%%%%%%
\begin{figure}
\includegraphics[width=\linewidth]{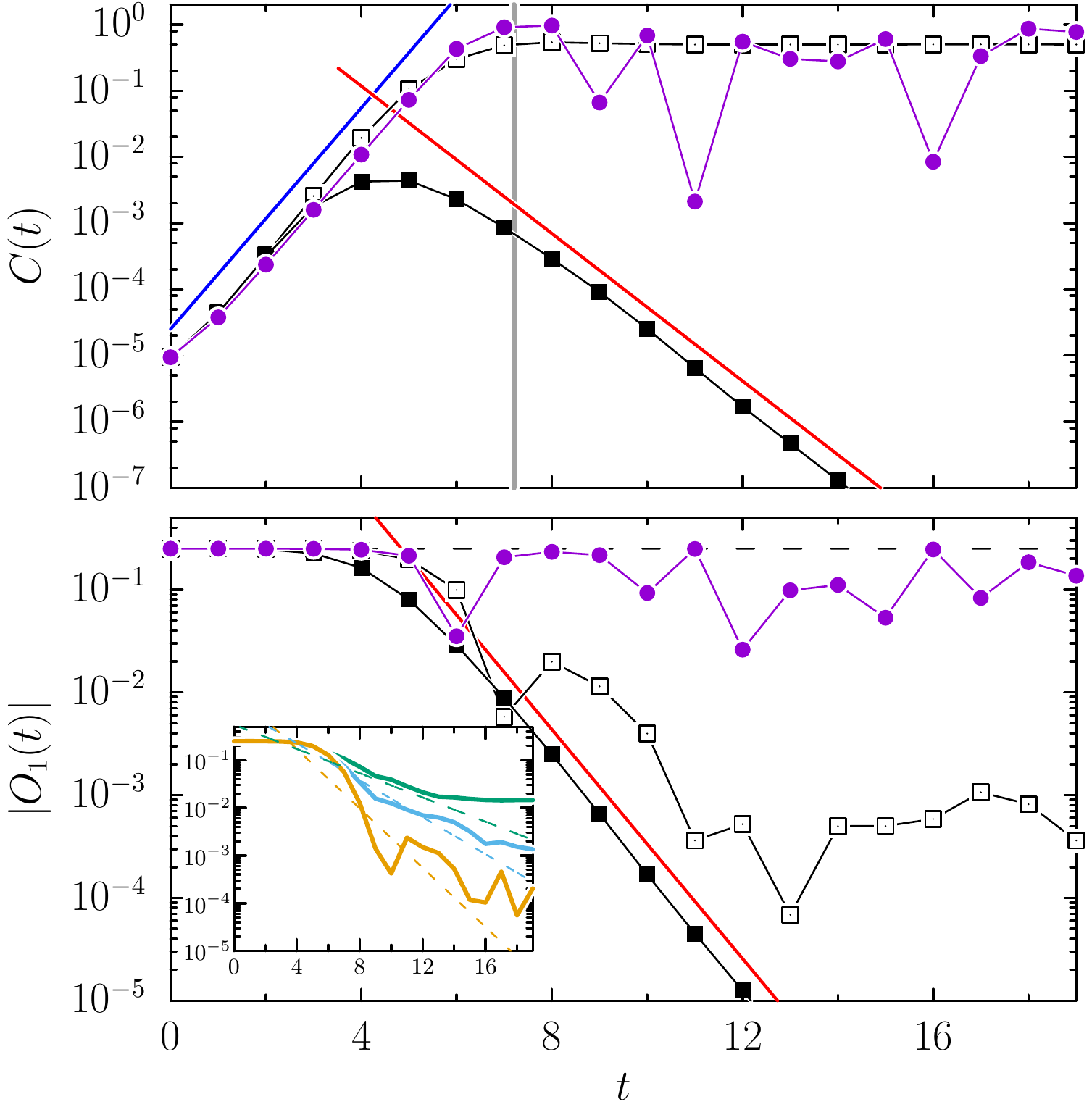}
\caption{
OTOC, \Eref{otocdef1}, (top panel) and $|O_1(t)|$, \Eref{eqO1}, (bottom panel) in the unitary map of \Eref{qcatmap}, for $k = 0$ ({\large {\color{miviol}$\bullet$}}) and $k = 0.02\ (\square)$, as well as in the dissipative (coarse-grained) map, defined in \Eref{coarse}, for $k = 0.02$ and $\epsilon=0.01\ (\blacksquare)$. The dashed horizontal line of the lower panel corresponds to the value $O_2=1/4$ obtained for the unitary map. The nonperturbed unitary results ($k=0$, $\epsilon=0$) are obtained analytically, the rest of the data comes from numerical calculations. For the latter, $N = 1024$ has been chosen. In the upper panel the vertical gray line marks the Ehrenfest time $t_{\rm E}$ and the blue line indicates $\exp(2\lambda_{\rm L} t)$, with $\lambda_{\rm L} = \ln [(3 + 5)/2]$. 
The red line on both panels shows a decay $|\alpha_1|^{2t}$, with $|\alpha_1|\approx 0.526$. The integer values of t stand for the number of iterations of the map. The lines joining the symbols are guides to the eyes. 
Inset: Additional data in the perturbed unitary map ($k \ne 0$, $\epsilon=0$) for $k = 0.325,\, 0.275,\, 0.25$ (solid lines from top to bottom), together with the decay $|\alpha_1|^{2t}$ corresponding to
$|\alpha_1|=0.864,\, 0.822,\, 0.698$ (dashed lines).  \label{figotoc}
}
\end{figure}
%%%%%%%%%%%%%%%%%%%%%%%%%%%%%%%%%%%%%%%%%%%%%%%%%%%%%%%%%%
%% 
%%% 
In Fig.~\ref{figotoc} (top) we compute $C(t)$ for the map of \Eref{qcatmap}, for $k=0$ and $k=0.02$. The case $k=0$ corresponds to the analytical derivation which we show below. There, the exponential 
growth  $C(t)\sim \exp[2\lambda_{\rm  L} t]$ for short times --  up to the Ehrenfest time \mbox{$t_{\rm E}=\ln(N)/\lambda_{\rm L}$} -- can be seen  explicitly. For $k>0$ the same Lyapunov behavior is observed. However, at the Ehrenfest time  the behavior is different. While for $k=0$ it oscillates (see below), for $k>0$ it saturates to $1/2$. 
The saturation of $C(t)$ follows from the unitary evolution in a finite-size system \cite{hashimoto2017,cotler2017,rammensee2018,chen2018,Jalabert2018}
and is set by the constant value of $O_2(t)$.

In the bottom of Fig.~\ref{figotoc} we show the behavior of $|O_1(t)|$. For the cases discussed above, $O_2(t)$ is trivially constant (dashed horizontal line).  
Interestingly, we notice that $O_1(t)\approx O_2(t)=1/4$ up to the Ehrenfest time, so that 
the exponential growth of $C(t)$ results from the difference between two  
quantities that are initially very close. 
Another remarkable observation is that for $t\approx t_{\rm E}$, $|O_1(t)|$ starts to decay, while $C(t)$  approaches a stationary value. 
The decay of $O_1(t)$ is exponential, and its rate depends on $k$. In the inset of Fig.\ref{figotoc} we present data for different values of $k$. The variation of the decay rate follows closely that of the RPR.
$O_1(t)$ is a correlation function and its long time decay rate for different values of $k$ appears generic in the numerical calculations. We show below that indeed the decay is governed by the RPR.
This fact will be further investigated below using a coarse-grained evolution.
%%%%
Thus, Fig.~\ref{figotoc} shows a summary of the main results of this Letter. In what follows, we 
provide an  analytical derivation of the Lyapunov regime for the systems we have considered.
We then introduce 
a coarse-graining propagator in order to gain access  to
the Ruelle-Pollicott regime.

\noindent
\textit{Lyapunov regime.--} %: 
We  proceed to derive an analytical expression for $C(t)$ in the case $k=0$. The
Weyl translation operators can be defined in terms of shift operators [\Eref{shift}]
as
\begin{equation}
\hT_\xi=\hat{V}^{\xi_q}\hat{U}^{\xi_p} \tau^{\xi_q\xi_p}
\end{equation}
with $\xi=(\xi_q,\xi_p) \in \mathbb{Z}_2$. They have the following properties
\begin{equation}
\label{properties}
\hT_\xi\hT_\chi = \tau^{<\xi,\chi>} \hT_{\xi+\chi}    \ ; \ \  
 \left[\hat{T}_\xi,\hat{T}_\chi \right] = 2i\sin\left(\frac{\pi}{N}<\xi,\chi> \right) \hT_{\xi+\chi},
\end{equation}
where $<\ ,\ >$ represents the symplectic product.
The important property of the translation operators is that, 
for quantum linear maps like 
$\hat{U}_M$
that quantize a symplectic linear transformation 
$M=\begin{pmatrix}
a&b\\
c&d
\end{pmatrix}$ with ${\rm det}[M]=1$, they transform ``classically'' as
\begin{equation}
\label{cattrans}
\hT_{M^t \xi}=\hat{U}^{\dagger t}_M\hT_\xi \hat{U}^{t}_M.
\end{equation}
This is the reason why the quantization of linear maps does not scramble operators, as it was pointed out in Ref.~\cite{chen2018}.
Now we define the following Hermitian operators
\begin{equation}
\hF_\xi =\frac{\hT_\xi-\hT^\dagger_\xi}{2i},
\end{equation}
which will allow us to define a family of OTOCs by replacing $\hA$ and $\hB$ in \Eref{otocdef1}. 
After a simple  calculation it is easy to see that these  OTOCs are given by
\begin{equation}
C(t)=-\frac{1}{N}\tr \left([\hF_\xi(t),\hF_\chi]^2\right)= \sin^2 (\frac{\pi}{N}<M^t\xi,\chi>),
\end{equation}
where we have defined
\begin{equation}
M^t\equiv 
\begin{pmatrix}
a_t & b_t \\
c_t & d_t
\end{pmatrix},
\end{equation}
where $a_t,\, b_t,\, c_t,\, d_t$ being integers modulo $N$ that 
grow exponentially with $\lambda_{\rm L}$, the logarithm of the largest
eigenvalue of $M$. This will happen until they become of order $N$ and mod$(N)$
kicks in.
From \Eref{XP} we see that $<M^t\xi,\chi >=-a_t$ for $\hX=\hF_{(1,0)}$ and $\hP=\hF_{(0,1)}$, and so we finally arrive at the following exact expression for the OTOC for $k=0$, 
\begin{equation}
\label{otoc_analyt}
C(t)=\sin^2\left(\frac{\pi a_t}{N}\right).
\end{equation}
Replacing $a_t=e^{\lambda_{\rm L} t}$ for early times such that $a_t<N$ 
\begin{equation}
C(t)\approx \left(\frac{\pi a_t}{N}\right)^2 = \frac{\pi^2}{N^2}e^{2\lambda_{\rm L} t},
\end{equation}
with $\lambda_{\rm L}$ the classical Lyapunov exponent of $M$.

This is the first important result of this Letter. 
We have  analytically shown that for short times the OTOC for a chaotic map, a paradigmatic example of quantum chaos, grows exponentially with a rate given by twice the classical Lyapunov exponent. 
Expanding the squared commutator and using cyclic properties of the trace and simple trigonometry, we have $O_2(t)=1/4$ and $O_1(t)=(1/4)\cos(2\pi a_t/N)$. 

\noindent
\textit{Ruelle-Pollicott regime.-- }
In what follows we explain the long time behavior observed in the case $k>0$. It is characterized by  
saturation of $C(t)$, which is in turn explained by the 
decay of $|O_1(t)|$. As stated before, this decay is given by 
$|\alpha_1|^{2 t}$, where $\alpha_1$ is the largest RPR, smaller than $\alpha_0=1$.
The RPRs are  the isolated eigenvalues $\{\alpha_i\}$ of the Koopman operator acting on a functional (Banach) space, less restrictive than ${\cal L}^2$, i.e. allowing some distributions \cite{Blank2002,Nonnenmacher2003}. They are located inside the unit circle and beyond some characteristic radius $r$. Therefore, when projected to a space orthogonal to the invariant density,
which  corresponds to $\alpha_0=1$, 
correlations decay asymptotically as $\alpha_1^t$ (if $|\alpha_1|>|\alpha_2|>\ldots$). We expand this discussion in the Supplemental Material.

The deep connection between the quantum propagator and the RPRs was established as a type of spectral quantum-classical correspondence, by introducing a coarse-grained propagator  \cite{Nonnenmacher2003,GarciaMata2003,GarciaMata2004,garciama2005}.
Such a propagator can be defined as a two-step superoperator 
\begin{eqnarray}
\label{coarse}
\hat{A}_{t+1}&=&\textbf{\textsf{D}}_\epsilon \left(\hat{U}^\dagger\hat{A}_t \hat{U}\right)\nonumber \\
&\stackrel{\rm def}{=}&\sum_\xi c_\epsilon(\xi) \hT^\dagger_\xi \hat{U}^\dagger\hat{A}_t \hat{U} \hT_\xi.
\end{eqnarray}
At each (discrete) time $t$ the unitary map $\hU$ is followed by an incoherent sum of all the possible translations in phase space with a (quasi-)Gaussian weight $c_\epsilon(\xi)$ centered at $\xi=0$. For convenience (see \cite{GarciaMata2004}) $c_\epsilon(\xi)$ is defined as the two-dimensional discrete Fourier transform of $\tilde{c}(\mu,\nu)=e^{-(\epsilon N/\pi)(\sin^2(\pi\mu/N)+\sin^2(\pi\nu/N))/2}$. It is approximately Gaussian with width proportional to $ 1/\epsilon$, 
and  $\epsilon$ therefore characterizing the size of the coarse graining.
The coarse graining $\textbf{\textsf{D}}_\epsilon$ introduces decoherence by dephasing noise. It can also be interpreted as an average over many different initial conditions, with Gaussian weight. The resulting propagator is nonunitary, and it is unital --it is a convex sum of unitary operators. 
One direct consequence of the nonunitarity is that $O^{(\epsilon)}_2(t)$ will now decay, and as a consequence, so will $C^{(\epsilon)}(t)$.

%%%%%%%%%%%%%%%%%%%%%%%%%%%%%%%%%%%%%%%%%%%%%%%%%%%%%%%%
\begin{figure}
\includegraphics[width=\linewidth]{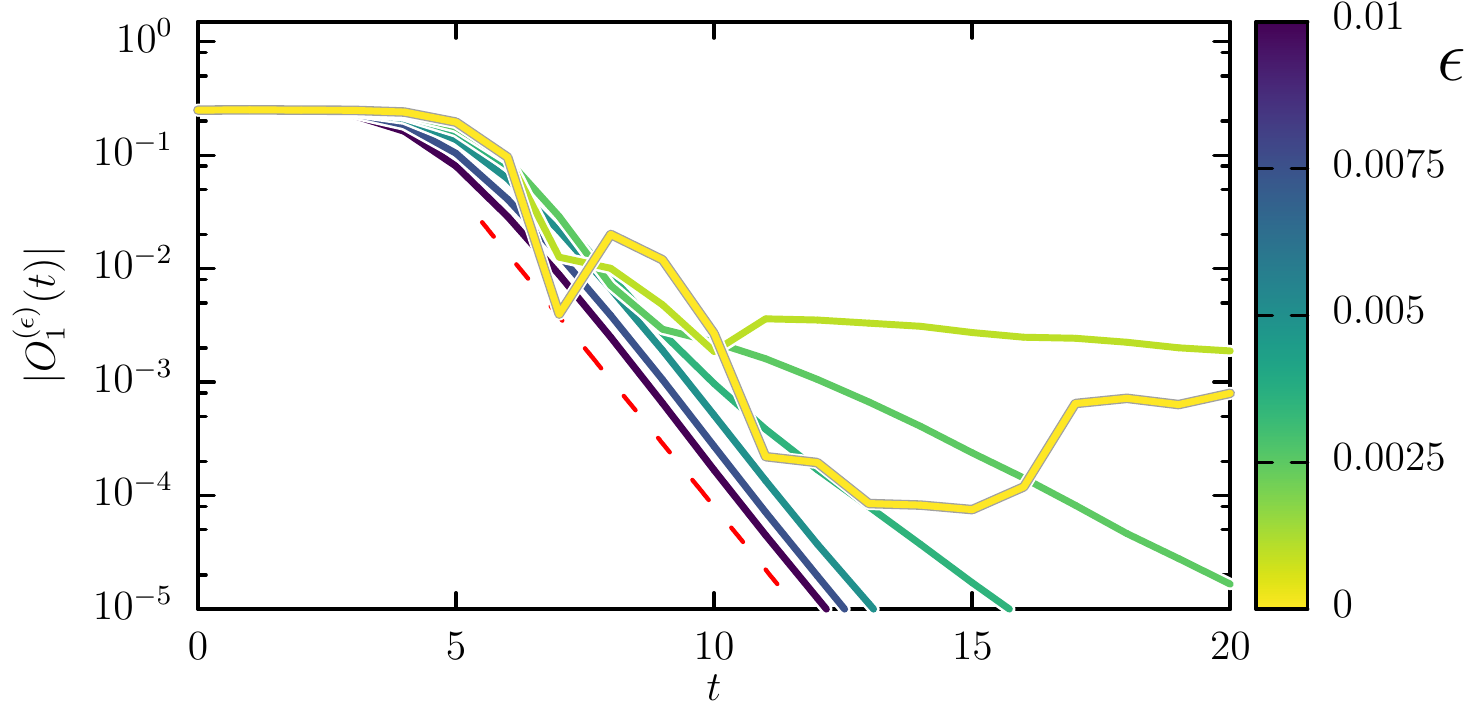}
\caption{$|O_1^{(\epsilon)}(t)|$ for different values of coarse-graining strength. Colors (shades) encode the values of $\epsilon$. Perturbed cat map with $k=0.02$, $N=1000$, $|\alpha_1|\approx 0.526$. 
The dashed (red) line indicates the Ruelle decay $|\alpha_1|^{2 t}$. 
\label{figdif}}
\end{figure}
%%%%%%%%%%%%%%%%%%%%%%%%%%%%%%%%%%%%%%%%%%%
It is proven in Ref. \cite{Nonnenmacher2003} that taking the limits $N\to\infty$ and $\epsilon\to 0$ in the appropriate order, the spectrum converges to that of an equivalent classical coarse-grained propagator whose spectrum in turn converges to the RPR. For completely chaotic systems there is a clear gap $|\alpha_1|<1$, and so for large times $O_i$ should decay as $|\alpha_1|^{2 t}$ (see Fig.~\ref{figotoc} top and bottom panels). Technically, obtaining the RPRs in the limit $N\to\infty$ becomes very hard, because it involves $N^2\times N^2$ matrices. Classical Ulam partitioning is not any easier. Rather, we profit from the fact that only the largest (in modulus) eigenvalue
is needed and an iterative method based on Lanczos power iteration can be used\cite{GarciaMata2004,Blum2000,Florido2002}. 
It essentially consists of iterating an initial arbitrary state forward and back,
building a matrix and solving a generalized eigenvalue problem. The key issue is that the existence of a finite gap reduces the size of the eigenvalue problem to solve drastically.  
The method guarantees that, in a limited region of $\epsilon$ and $N$ ($\epsilon N=$ const \cite{Nonnenmacher2003,garciama2005}), we can
extract from the unitary operator a certain number of resonances that are
independent of both $\epsilon$ and $N$ so that they are classical ($N\to\infty$)
characteristics of the map. Numerically for $N\approx 1000$ we can extract up to around 10 -- 15
of the largest resonances.
As an alternative, since $c_\epsilon(\xi)$ is quasi-Gaussian, it effectively truncates the propagator in Fourier space, and this fact can be used to compute the spectrum in 
what is sometimes called the chord representation \cite{aolita2004}.

In Figs.~\ref{figotoc} and \ref{figdif} the effect of the coarse graining is shown for both $C(t)$  and $|O_1(t)|$. Since $O^{(\epsilon)}_2(t)$ is no longer constant (see Supp. mat.), $C^{(\epsilon)}(t)$ does not saturate, but decays also according to the largest RPR. 
In Fig.~\ref{figdif} we show $|O_1^{(\epsilon)}(t)|$ for different values of $\epsilon$. We can see how the Ruelle-Pollicott behavior, even though already present without coarse graining, is progressively unveiled and lasts longer as $\epsilon$ becomes larger. Eventually, after some threshold value the decay of $|O_1|$ saturates to  $|\alpha_1|^{2 t}$ (dashed red line). This behavior is expected to be valid for a limited range of $\epsilon$. For large enough values, the effect of $\su{D}_\epsilon$ will dominate over the unitary dynamics.

We point out that the decay governed by RPR's presented by the OTOC, beyond the Ehrenfest time, for dissipative time evolution is reminiscent of the long time behavior exhibited by the Loschmidt echo   \cite{GarciaMata2003,GarciaMata2004,garciama2005}. Such concomitance suggests a link between both time-dependent correlators \cite{zangara2016,kurchan2016}.
In the Supplemental Material  we present numerical results for two other examples of chaotic maps. The standard  and Harper maps. The numerics for these other maps fully support the results presented here. 

\noindent
\textit{Conclusions.--}
The OTOC is a powerful tool to characterize chaos in a great variety of domains from single particle to many-body physics, up to black holes and high energy thermodynamics. Part of this power comes from the possibility to relate its time dependence to classical quantities like the Lyapunov exponent, that can be independently determined. We have analytically shown, for simple systems, how the classical Lyapunov exponent appears explicitly in the early time dependence of the OTOC.  Moreover, we have also shown 
that   after the Ehrenfest time the approach to saturation of the OTOC or its eventual decay is determined by the RPRs, which  are the classical quantities responsible for the decay of correlations in strongly chaotic systems.
This leads us to conclude that the main traits of classical chaos are embedded in 
both the short and long time dynamics of the OTOC.

We thank H. M. Pastawski for helpful discussions.
The authors have received funding from CONICET (Grant No PIP 11220150100493CO)
ANPCyT (PICT-2016-1056 and  PICT 2014-3711), UBACyT (Grant No 20020130100406BA), and a  binational collaboration project funded by CONICET and CNRS (PICS No. 06687)
%%%%%%%%%%%%%%%%%%%%%%%%%%%%%%%%%%%%%%%%%%%%%%%%%%%%%%%%%%%%%%%%%%%%%%%%%%%%%%%%
%\bibliographystyle{apsrev4-1}
%\bibliography{refs}
%%%%%%%%%%%%%%%%%%%%%%%%%%%%%%%%%%%%%%%%%%%%%%%%%%%%%%%%%%%%%%%%%%%%%%%%%%%%%%%% 
%% aca pegar .bbl
%\end{document}		%%%%  		***				FIN				***			%%%%
%%%%%%%%%%%%%%%%%%%%%%%%%%%%%%%%%%%%%%%%%%%%%%%%%%%%%%%%%%%%%%%%%%%%%%%%%%%%%%%%
\onecolumngrid
\newpage
\appendix
\begin{center}
{\large\textbf{Supplemental material to\\  
``Chaos signatures in the short and long time behavior of the out-of-time ordered correlator''
}}
\end{center}
%%%%%%%%%% Prefix a "S" to all equations, figures, tables and reset the counter %%%%%%%%%%
\setcounter{equation}{0}
\setcounter{figure}{0}
\setcounter{table}{0}
\setcounter{page}{1}
\makeatletter
\renewcommand{\theequation}{S\arabic{equation}}
\renewcommand{\thefigure}{S\arabic{figure}}
\renewcommand{\bibnumfmt}[1]{[S#1]}
\renewcommand{\citenumfont}[1]{S#1}
\label{suppmat}
%\appendix
%\section*{SUPPLEMENTARY MATERIAL}
%\label{appendix}
\subsection*{Long time decay of quantum chaotic coarse-grained dynamics }
We have defined the coarse grained superoperator as a map between operators, consisting of a unitary transformation followed by a convex sum of translations in phase space 
\begin{equation}
\se (\hat{O})\stackrel{\rm def}{=}\sum_\xi \hT_\xi^\dagger \hU^\dagger \hat{O} \hU \hT_\xi,
\end{equation}
where $\hU$ is a unitary map in a 2-torus phase space, $\hT_\xi$ are displacements by a vector $\xi \in \mathbb{Z}_2$, and $\hat{O}$ some arbitrary operator. Since $\se$ is a convex sum of unitaries it is completely positive and contracting. Moreover, it is unital, i.e. it has one eigenvalue 1, corresponding to the uniform density $\hat{\rho}=\hat{I}/N$. The rest of the spectrum lies inside the unit circle. Furthermore, it is not normal, so left and right eigenoperators are in general different
\begin{equation}
\se \hat{R}_i=\alpha_i\hat{R}_i\ ,\ \  \se^\dagger\hat{L}_i=\alpha_i^*\hat{L}_i.
\end{equation}
The left and right eigenoperators $\hat{R}_i$, $\hat{L}_i$ form a biorthogonal set (with the inner product defined by the trace)
\begin{equation}
\tr(\hat{L}_i^\dagger\hat{R}_i)=\delta_{ij}.
\end{equation}
In addition, we assume the eigenoperators to be normalized as 
$\tr(\hat{L}_i^\dagger\hat{L}_i)=\tr(\hat{R}_i^\dagger\hat{R}_i)=1$, and their eigenvalues 
ordered by decreasing modulus $1>|\alpha_1|\ge |\alpha_2|\ge \ldots |\alpha_{N^2-1}|$. For simplicity
we assume the spectrum to be non-degenerate. For $\alpha_0=1$ we have 
$\hat{R}_0=\hat{L}_0=\hat{I}/N$.
Then, the spectral decomposition looks like
\begin{equation}
\label{spectra}
\se=\sum_i \alpha_i\hat{R}_i\tr(\hat{L}_i^\dagger\ . \ ),
\end{equation}
where $i=0,\ldots,N^2-1$,
and the dot implies the action of $\se$ on an operator.
In Ref.~\cite{Nonnenmacher2003} it was demonstrated that this type of spectral decomposition is possible, and that taking suitable limits (first  $N\to \infty$ and then $\epsilon\to 0$) the eigenvalues correspond to the Ruelle-Pollictott resonances. 

Let us now consider the evolution of an operator, in particular  $\hX$, which can be expanded  
as
\begin{equation}
\hX=\sum_i x_i\hat{R}_i, \ \text{with}\ x_i=\tr(\hat{L}_i^\dagger \hX).
\end{equation}
Then, the evolution up to (discrete) time $t$  is given by
\begin{equation}
\hX(t)=\se^t\hX=\sum_i x_i \alpha_i^t \hat{R}_i.
\end{equation}
Additionally, we have $\tr(\hX)=x_0=0$ and thus, asymptotically, after the Ehrenfest time, 
we have 
\begin{equation}
\hX(t)\approx \alpha_1^{t} \hat{R}_1,
\end{equation}
and
\begin{equation}
|O_1^{(\epsilon)}(t)|=|\alpha_1|^{2t}\, |x_1| |\tr(\hat{R}_1\hP\hat{R}_1\hP)|.
\end{equation}
We have assumed  $\alpha_1$ to be real, which is usually the case. If it is complex, then 
oscillations are superposed with the exponential decay. We have also observed that if many $\alpha_i$ are close in modulus, then the decay is less clear, and an average behavior is more likely to be observed. 
An analogous analysis holds for $O_2(t)$, and 
consequently for $C(t)$ as well.
\subsection*{Examples for other maps}
In this supplementary part we discuss two other maps beyond that of the main text, in order to test the universality of our results.
Both of these maps can be derived from  kicked Hamiltonians 
The first one is the Chirikov standard map \cite{DimaScholar}
\begin{equation}
\begin{array}{lll}
p'&=&p+\frac{K}{2\pi}\sin(2 \pi q)\\
q'&=&q+p'.
\end{array}\ \ \mod 1
\end{equation}
which is the Poincar\'e section of a kicked rotator.
Below a certain critical value $K_c\approx 0.971635\ldots$ \cite{Greene1979}, 
the motion of the standard map in momentum is 
limited by KAM curves. Above $K_c$ there is unbounded motion  in $p$. For 
very large $K$, there are no significant islands and the motion is essentially chaotic.
%%%%%%%%%%%%%
\begin{figure}[h]
\includegraphics[width=0.9\linewidth]{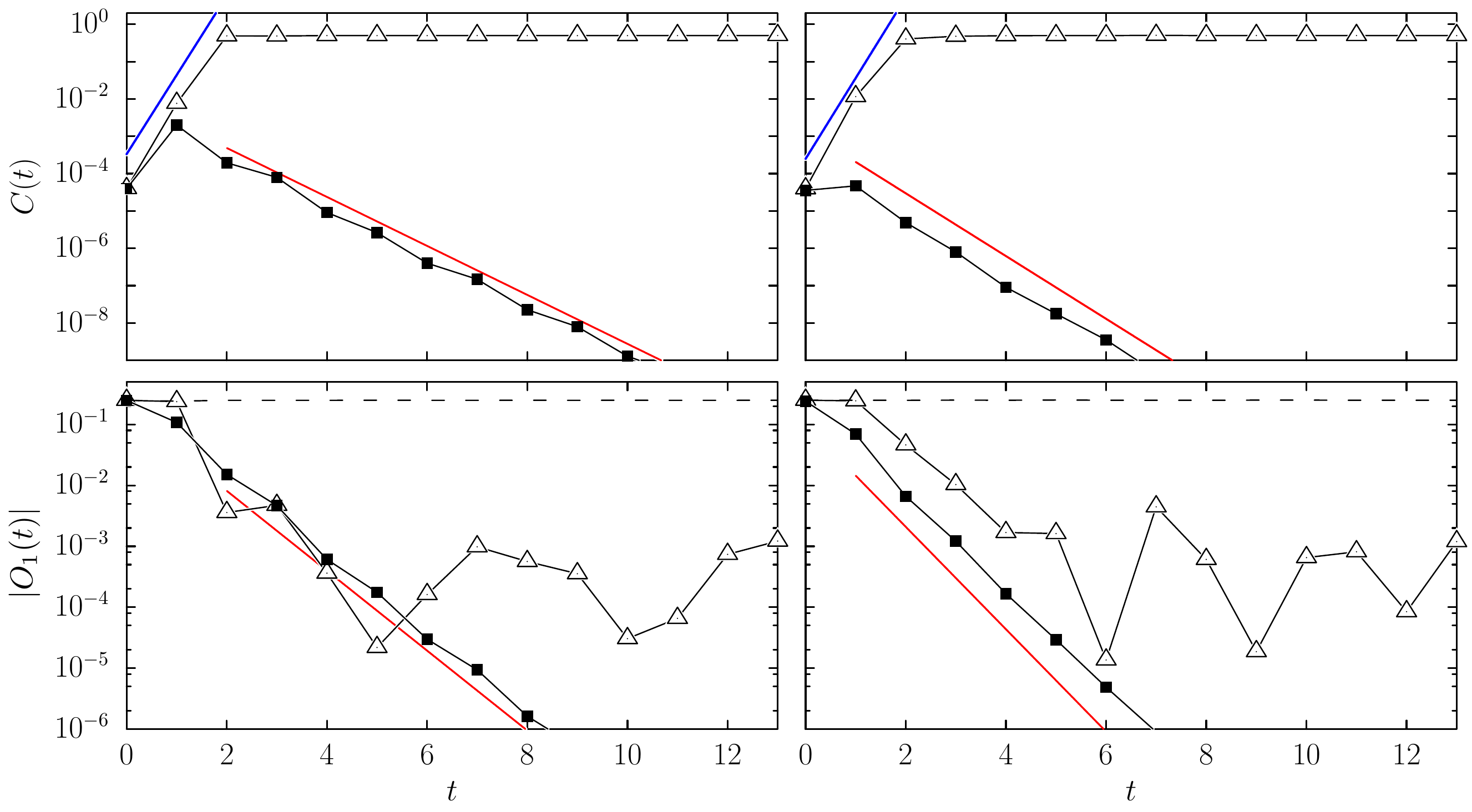}
\caption{(top) $C(t)$ (bottom) $|O_1(t)|$ for (left) the standard map with $K=19.74$ and $|\alpha_1|\approx 0.47$;  (right) the Harper map with $K=0.94$ and $|\alpha_1|\approx 0.38$. The blue lines correspond to $e^{2\lambda_{\rm L}t}$ and the red ones to $|\alpha_1|^{2 t}$.
The behavior for these two maps is equivalent to the one observed in Fig.~1 of the main text for the cat map.
\label{figotoc_supp}}
\end{figure}
%%%%%%%%%%%%%%%%%%%%%%%

The second map studied in this section is
the Harper map, which
is an approximation of the motion of a kicked charge under the action of an external magnetic field \cite{Dana1995,ArtusoScholar}. 
\begin{equation}
\begin{array}{lll}
p'&=&p-K_1\sin(2 \pi q)\\
q'&=&q+K_2\sin(2 \pi p')
\end{array}\ \ \mod 1.
\end{equation}
For simplicity we
only consider the symmetric case $K_1=K_2= K$.
In the case of the Harper map, for $K<0.11$, the dynamics described by the associated classical map is regular, while for $K>0.63$ there are no remaining visible regular islands \cite{leboeuf1990}.

The advantage of using these types of maps is that the quantum version takes the simple form  \cite{DimaScholar,leboeuf1990}
%%%%%%%%%%%%
\begin{equation}
\label{qmap}
\hU_M=\hU(p)\hV( q)\ \left\{
\begin{array}{lll}
U_{K}^{({\rm Standard})}&=& e^{-i\pi \frac{p^2}{N}}e^{-i 2\pi NK\cos(2 \pi q/N) } \\
 & &\\
U^{({\rm Harper})}_{K}&=& e^{i 2 \pi N K \cos(2 \pi p/N)}e^{i 2\pi N K \cos(2 \pi q/N) } 
\end{array}
\right. ,
\end{equation}
were $q,p=0,1,\ldots,N-1$.
Both maps in Eq.~\ref{qmap} are efficient to implement numerically using fast Fourier transforms.

\begin{figure}[h]
\includegraphics[width=0.9\linewidth]{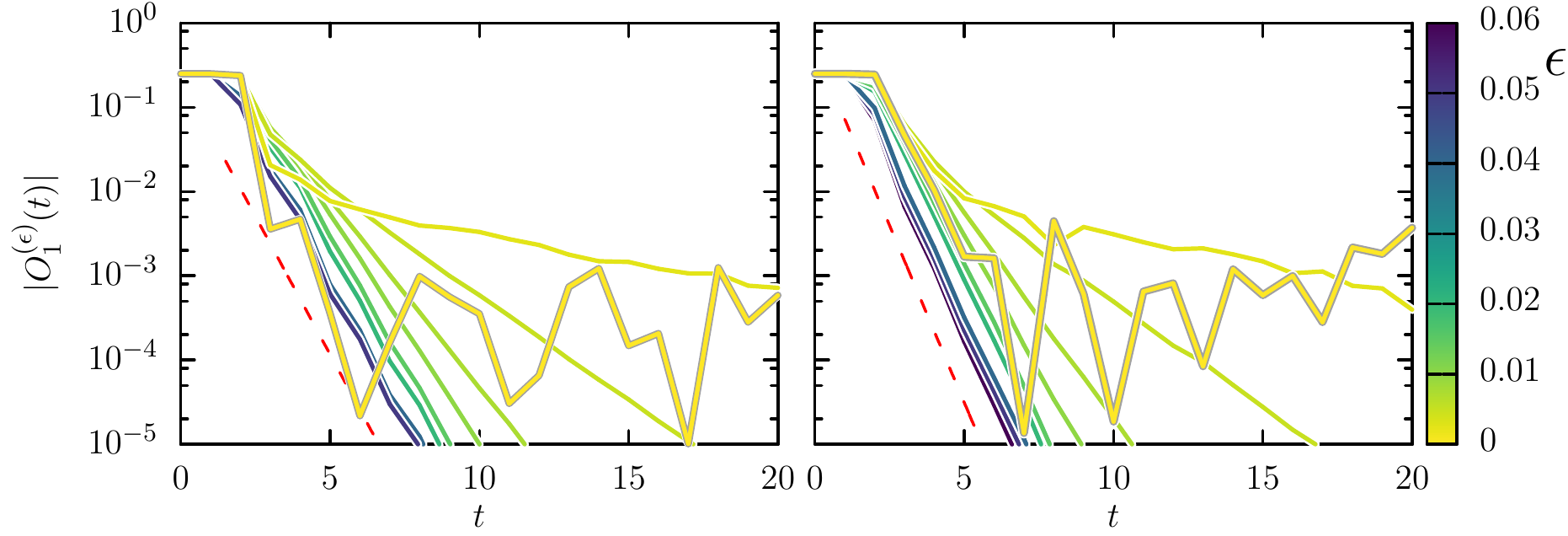}
\caption{$|O_1^{(\epsilon)}(t)|$ for different values of coarse-graining strength ($d=500$). Colors (shades) encode the values of $\epsilon$. (left)  Standard map with $K=19.74$, $|\alpha_1|\approx 0.47$; (right) Harper map with $K=0.94$, $|\alpha_1|\approx 0.38$.  
The dashed (red) line indicates the Ruelle decay $|\alpha_1|^{2t}$. The behavior obtained for these two maps is similar to that of the cat maps presented in Fig.~2 of the main text.
\label{figdif_supp}}
\end{figure}

In  Figs.~\ref{figotoc_supp} and \ref{figdif_supp} we present results that 
support and extend those presented in the main text for the cat map.  
For linear maps like the
(unperturbed) cat map of the main section, the Lyapunov exponent $\lambda=\ln \alpha$, where $\alpha$ is the largest eigenvalue of the monodromy matrix.  On the other hand, for nonlinear maps 
such a quantity depends on the initial condition 
and an average should be made. 
The standard definition is $\lambda=\langle\lim_{t\to\infty}\lim_{d(0)\to 0}(1/t)\ln(d_{X_0}(t)/d_{X_0}(0))\rangle_{X_0}$, where $d_{X_0}(t)$ represents a distance in phase space and the bracket represent a phase space average over the initial condition $X_0$. In \cite{Rozenbaum2017} a generalized Lyapunov exponent $\Lambda$ was defined where the average is done before computing the logarithm. 
In the computations shown in Fig.~\ref{figdif_supp} the difference between $\lambda$ and $\Lambda$ is imperceptible.
The subsequent approach to saturation given by RPR, is very convincing (Fig.~\ref{figotoc_supp}). In Fig.~\ref{figdif_supp}
the dependence on the coarse-graining parameter $\epsilon$ is exposed. For a range of relatively large values of $\epsilon$ the decay rate saturates. However if $\epsilon$ becomes too large then the coarse-graining dominates the dynamics and no trace of the original classical map is expected to appear.
%%%%%%%%%%%%%%%%%%%%%%%%%%%%%%%%%%%%%%%%%%%%%%%%%%%%%%%%%%%%%%%%%%%%%%%%%%%%%%%
%\bibliographystyle{apsrev4-1}
%\bibliography{refs}
%%%%%%%%%%%%%%%%%%%%%%%%%%%%%%%%%%%%%%%%%%%%%%%%%%%%%%%%%%%%%%%%%%%%%%%%%%%%%%%
%% aca pegar .bbl
\end{document}		%%%%  		***				FIN				***			%%%%
%%%%%%%%%%%%%%%%%%%%%%%%%%%%%%%%%%%%%%%%%%%%%%%%%%%%%%%%%%%%%%%%%%%%%%%%%%%%%%%%